\documentclass[a4paper, oneside, twocolumn, notitlepage, 10pt]{extarticle_ecoc2015}
\usepackage{ecoc2015}
\usepackage{dblfloatfix}

\usepackage{pgfplots}
\usetikzlibrary{backgrounds,intersections,positioning}
\usepackage{bm}
\usepackage{alucolor}
\usepackage{comment}

\newcommand\blfootnote[1]{%
  \begingroup
  \renewcommand\thefootnote{}\footnote{#1}%
  \addtocounter{footnote}{-1}%
  \endgroup
}
\begin{document}
\selectlanguage{american}    


\title{Distributed Transmission and Spatially Coupled Forward Error Correction in Regenerative Multipoint-to-Point Networks\vspace*{-1ex}}%


\author{
    Laurent Schmalen,\textsuperscript{(1)} Tobias A. Eriksson,\textsuperscript{(1)}
    Fred Buchali,\textsuperscript{(1)} Roman Dischler,\textsuperscript{(1)} and Ulrich Gebhard\textsuperscript{(1)}
\vspace*{-1ex}}

\maketitle                  


\begin{strip}
 \begin{author_descr}

   \textsuperscript{(1)} Nokia Bell Labs, Lorenzstr. 10, Stuttgart, Germany,
   \uline{\{firstname.lastname\}@nokia-bell-labs.com}\vspace*{-1ex}
 \end{author_descr}
\end{strip}

\setstretch{1.08}


\begin{strip}
  \begin{ecoc_abstract}
    We investigate the performance of coded modulation for multi-hop regenerative optical networks. We analyze options for computing decoder input LLRs, show reach increases by optimized regenerator placement and experimentally compare strategies and guidelines for distributed FEC.\vspace*{-1ex}
  \end{ecoc_abstract}
\end{strip}


\section{Introduction}
Future optical metro networks require novel network architectures tailored to Ethernet transport to cope with the increasing demand for flexible bit rates and traffic flows as required by, e.g., dynamic data center interconnects. Traditional transport networks like SDH or OTN encapsulate Ethernet to add the forward error correction (FEC) and monitoring needed for long-haul transport which Ethernet does not provide. Their point-to-point connections, however, are provisioned and static, they cannot follow dynamic traffic demand.\blfootnote{This work is supported by the German BMBF in the scope of the CELTIC+ project SENDATE-TANDEM.}

\begin{figure}[htb!]
\includegraphics[width=\columnwidth]{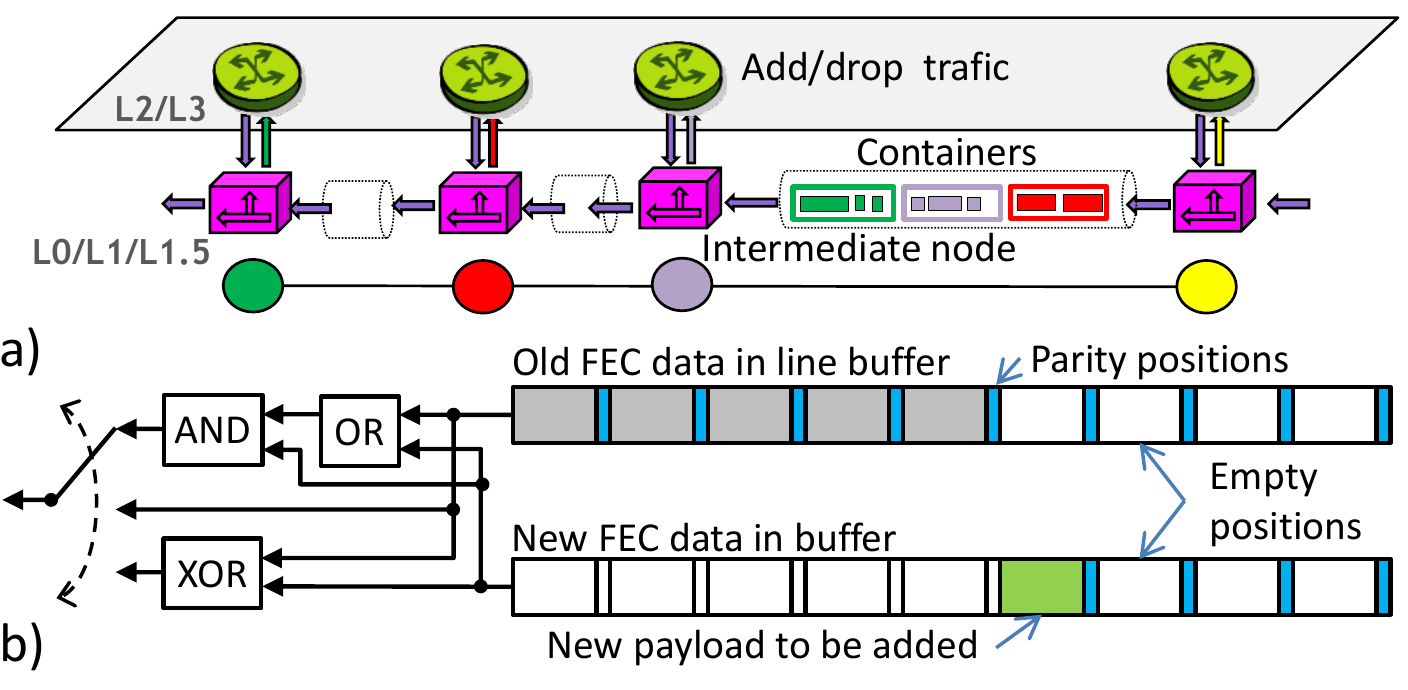}
\caption{a) Architecture with switching integrated into the physical layer and b) generation of new FEC portion by XOR of old FEC parity bits and by replacement of information bits.}
\label{fig:network_arch}
\end{figure}

In contrast, we enable dynamic transport connections by aggregating Ethernet packets into fixed-length transport containers, which are individually routable based on a very short container header~\cite{refDischlerECOC16}. The container capacity can be shared by multiple sources along its route, resulting in a distributed multipoint-to-point (DMPP) connection scheme. The network architecture is illustrated in Fig.~\ref{fig:network_arch}-a) and described in detail in~\cite{refOE}. The node architecture is based on the digital regenerator architecture experimentally investigated in~\cite{refBuchaliECOC15}, which provides signal conditioning without costly FEC decoding in transit nodes. Fast, distributed encoding of transit containers is achieved by a FEC design described in this paper.

The container payload consists of information bits followed by periodically introduced parity bits. At an intermediate node, the new data to be added is written into a free portion of the container (using AND-OR; the occupancy of the container is indicated in the container header) and the parity bits are updated by XOR as shown in Fig.~\ref{fig:network_arch}-b).

In this paper, we investigate soft-decision decodable codes for DMPP networking. We focus on spatially coupled (SC) LDPC codes due their excellent performance and potential record net coding gains (NCGs)\cite{refAmirSchmalen}. We first investigate potential gains using soft decision decoding and regeneration, experimentally show reach increases by regeneration  with 32GBaud PM-16QAM and PM-32QAM and finally also show how SC-LDPC codes shall be designed for this application.\vspace*{-0.6ex}

\section{Exemplary Two-Hop Network}

\begin{figure}[b!]
\centering
\vspace*{-4ex}
\includegraphics[width=\columnwidth]{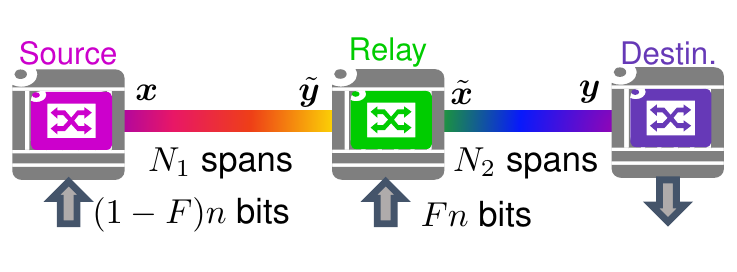}
\vspace*{-4ex}
\caption{Exemplary network with source, relay and destination.}
\vspace*{2ex}
\label{fig:example_network}
\end{figure}

\begin{figure*}[t!]
\vspace*{-1ex}
\includegraphics[width=\linewidth]{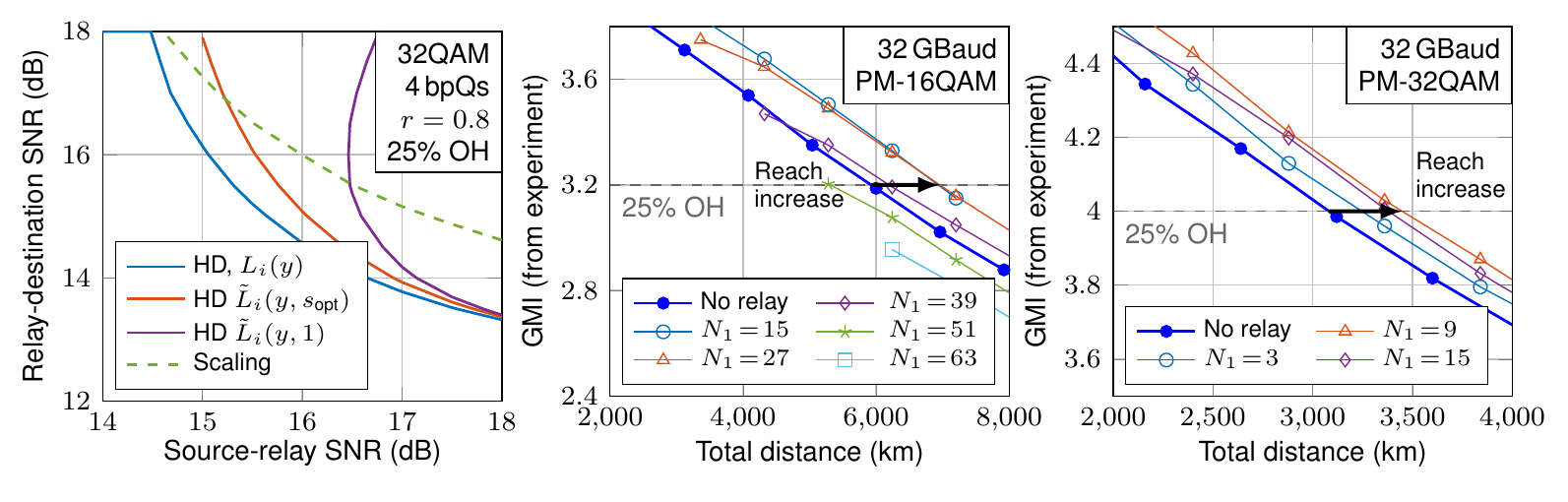}
\vspace*{-4ex}
\caption{Required SNR regions for HD relaying and different LLR computation rules and scaling at the relay (left). Experimentally measured results with an HD-relay placed after $N_1$ spans of 80\,km for 32\,GBaud PM-16QAM (middle) and PM-32QAM (right).}
\vspace*{-0.5ex}
\label{fig:resultGMI_noF}
\end{figure*}

To illustrate the concept, we consider the exemplary network shown in Fig.~\ref{fig:example_network}. A source node transmits a vector $\bm{x}$ consisting of $n$ modulation symbols over an optical link to an intermediate node, denoted \textit{relay}. The relay receives a noisy version $\tilde{\bm{y}}$ of $\bm{x}$ and retransmits a modified version $\tilde{\bm{x}} = f_{\text{Tx-DSP}}(f_{\text{R}}(f_{\text{Rx-DSP}}(\tilde{\bm{y}})))$, wherein $f_{\text{Rx-DSP}}(\cdot)$ describes the receiver DSP,   $f_{\text{Tx-DSP}}(\cdot)$ the transmit DSP (e.g., pulse-shaping, etc.) and $f_{\text{R}}(\cdot)$ a regenerating function. Here we mostly consider the  \textit{symbol hard-decision} (HD) regenerating function, i.e., $
f_{\text{R}}(x) = \arg\min_{\omega\in\Omega}\Vert x-t\Vert_2$, which outputs the constellation symbol $\omega\in\Omega$ minimizing the Euclidean distance to the received symbol. For comparison, we also consider the \textit{scaling} regenerating function $f_{\text{R}}(x) = \eta x$, i.e., which simply scales the input, e.g., to adjust the launch power for the second fiber segment. It has been shown\cite{refBuchaliECOC15,SorokinaRegen,refSchmalenImplicit} that the HD regenerating function can lead to significant reach increase compared to the scaling when multiple relays are equidistantly spaced along the link.

Here we consider a single relay that can be placed anywhere along the link and we allow basic network functionality based upon the OE~\cite{refOE} protocol and mechanisms: the relay node is allowed to add (and also drop, which is not considered in the example as it is a trivial extension of the add case) traffic that is directed towards the destination. From a block of $n$ symbols, the relay node adds $Fn$ information symbols, where $F$ is the \textit{relay loading factor}. The separately transmitted header\cite{refDischlerECOC16} describes the payload composition.

At the receiver, we need to take into account the fact that (a fraction $1-F$ of) the data is transmitted via an equivalent channel composed of the first segment including the regeneration and the second segment. The relay (or a chain of relays) is modeled by a single discrete memoryless channel with transition matrix $\bm{W}$, where $W_{i,j} = P(\tilde{x}=\varphi(i)|x=\varphi(j))$ and $\varphi(i) = \omega_i$ is the modulation map assigning a constellation symbol $\omega_i\in\Omega$ to an integer $i$. At the receiver employing soft-FEC decoding with bit-interleaved coded modulation (BICM), the first step is to compute log-likelihood ratios (LLRs). We subdivide the constellation $\Omega$ into two sets $\Omega_b^i$, which contain those symbols where the $i$-th bit of the bit patterns takes on the value $b\in\{0;1\}$. The LLR for bit-level $i$ is computed as\cite{refSchmalenImplicit}\vspace*{-2ex}
\[
L_i(y) \!=\! \ln\!\left(\!\frac{\sum\limits_{x\in\Omega_0^i}\sum\limits_{j=1}^M W_{j,\varphi^{-1}(x)}\exp\!\left(\!-\frac{\Vert y\!-\!\varphi(j)\Vert_2^2}{2\sigma_{H}^2}\!\right)}{\sum\limits_{x\in\Omega_1^i}\sum\limits_{j=1}^M W_{j,\varphi^{-1}(x)}\exp\!\left(\!-\frac{\Vert y\!-\!\varphi(j)\Vert_2^2}{2\sigma_{H}^2}\!\right)}\right)
\vspace*{-1ex}
\]
assuming Gaussian noise of variance $\sigma_{H}^2$ in the final hop, which is in accordance with the GN-model. $L_i(y)$ can be easily approximated by a piece-wise linear function of $y$. 

Sometimes, this LLR computation rule cannot be directly implemented, e.g., if only a legacy module with a conventional LLR calculator is available. In this case, we scale the conventional LLRs as $\tilde{L}_i(y,s) = s\cdot L_{\text{conv.}}(y)$,
where $s$ is a scaling factor chosen to maximize the GMI\cite{IvanovGMI} with $s =  \arg\inf_{\zeta\geq 0}\sum_{i=0}^m \mathbb{E}\{\log_2(1+e^{-\zeta (-1)^{b_i}\tilde{L}_i(y,1)})\}$,
where $b_i$ is the transmitted bit in bit level $i$.

To experimentally evaluate the method, we use a single recirculating loop with 80\,km spans of SSMF. As the targeted applications are metro networks and data-center interconnects, we use PM-32QAM and PM-16QAM with GMI-maximizing bit mappings. The 32~Gbaud PM-16QAM and PM-32QAM signals are generated using an 88\,GS/s four-channel DAC and a dual-polarization IQ-modulator. The signals are detected using a conventional polarization-diverse coherent receiver and digitized with a 33\,GHz BW, 80\,GS/s oscilloscope. We use a data-aided DSP scheme, where after HD regeneration, the symbols are re-uploaded to the DAC and re-transmitted over the loop to emulate the second hop.

First we consider the case $F=0$, i.e., the relay acts purely as regenerator. We first model the two optical links as AWGN channels. We are interested in the range of allowable SNRs between source-relay and relay-destination. We target FEC with $25$\% OH. In Fig.~\ref{fig:resultGMI_noF}, we show the SNR contours above which reliable transmission is possible. The HD relay outperforms the scaling relay and  the legacy LLR function with optimized scaling $s_{\text{opt}}$ can offer very good performance. In the middle and right plot of Fig.~\ref{fig:resultGMI_noF}, we show the experimental transmission performance. We can see that a single relay, if placed properly, can already lead to a reach increase of 10-17\,\%. The reach increase is larger for PM-16QAM due to the reduced implementation penalty at the relay. Note that the relay may also degrade performance if the source-relay distance becomes too large.

\begin{figure*}
\vspace*{-1ex}
\begin{tabular}{ccc}
\includegraphics{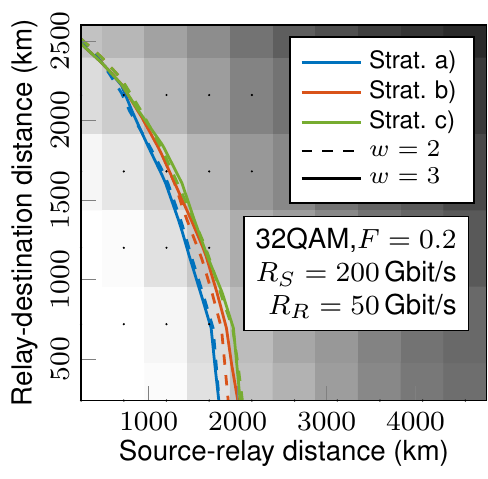} & \includegraphics{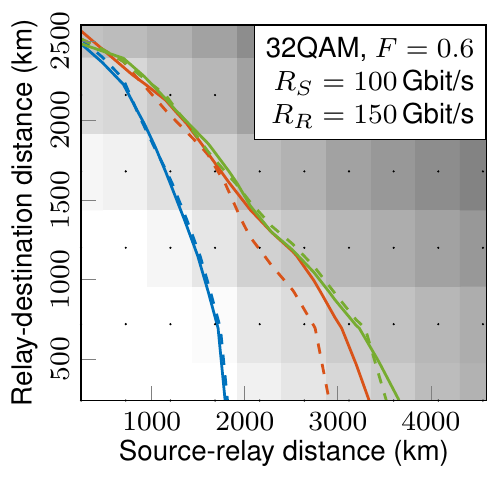} & \includegraphics{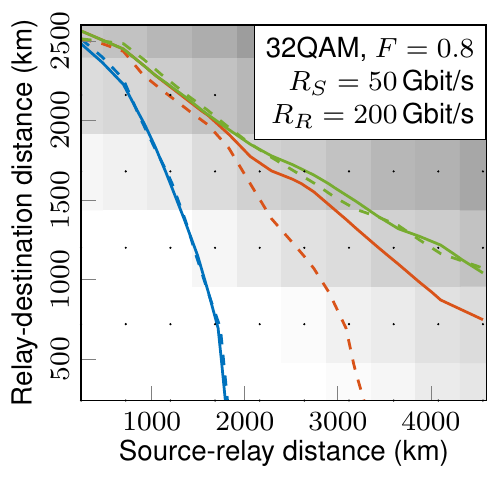}
\end{tabular}
\vspace*{-1ex}
\caption{Contour plots of achievable source-relay and relay-distance distance pairs for 32GBaud PM-32QAM (from experiment) for different loading factors $F$, three different coding strategies from Fig.~\ref{fig:coding_strategies} and two coupled code designs ($w\in\{2,3\}$). The shaded background shows the experimentally obtained values (at the centers of rectangles) for strategy c) with $w=3$.}\vspace*{0.8ex}
\label{fig:contour_results}
\end{figure*}

\begin{figure}[t!]
\centering
\vspace*{-1.5ex}
\includegraphics[width=\columnwidth]{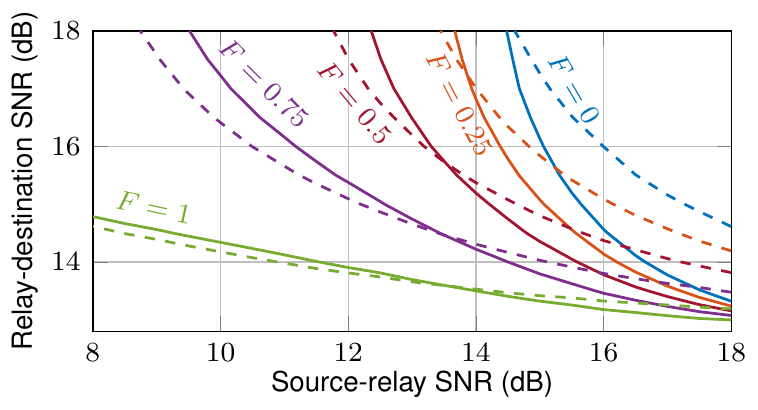}
\vspace*{-2.5ex}
\caption{SNR regions for different regenerators (HD: solid lines, scaling: dashed lines) and different loading factors $F$.}
\label{fig:resultGMI_withF}
\end{figure}

Next, we consider different loading factors $F$. We assume that the relay is able to carry out perfect interleaving. We compare HD regeneration with scaling. Fig.~\ref{fig:resultGMI_withF} shows the range of acceptable SNR pairs for the different setups. Obviously, when increasing $F$, the performance becomes less dependent on the SNR of the source-relay hop and the penalty from scaling becomes smaller. Note that the parity bits are always generated at the source and updated at the relay, hence they are always limited by the overall source-destination SNR.

\vspace*{-1ex}

\section{Coding Strategies and Experimental Results}
\begin{figure}[t!]
\includegraphics[width=\columnwidth]{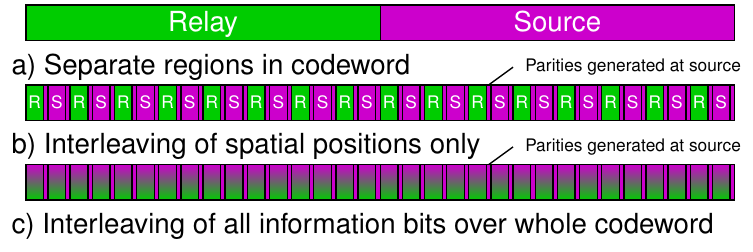}
\caption{Three different coding strategies used.}
\label{fig:coding_strategies}
\end{figure}

Next, we apply actual coding with a family of high-performing SC-LDPC codes\cite{refAmirSchmalen}. We compare unit-memory codes (coupling width $w=2$) with codes having syndrome former memory 2 ($w=3$). Due to linearity, we separately encode the new data at the relay, replace the systematic part of the codeword and add the parity-part on bit-level (see Fig.~\ref{fig:network_arch}-b)). We compare three different coding strategies shown in Fig.~\ref{fig:coding_strategies}. In strategy a), we use the first $FN$ bits of the codewords for the portion added in the relay. In strategy b), each of the $L=60$ spatial positions of the SC-LDPC code corresponds to either a source or relay data portion. We assign them such that always two spatial positions corresponding to a source portion are at maximum distance (according to an asymptotic optimization). Finally, in strategy c), we  interleave the bits equally over the codeword. The first two strategies can be easily mapped to the container/sub-container structure of the OE protocol\cite{refOE} and we consider strategy c) as baseline. Note that in strategies b) and c), the parities are all generated at the source due to the recursive nature of the code.

In Fig.~\ref{fig:contour_results}, we plot the contour lines highlighting the distance pairs above which decoding with a windowed decoder is possible for different $F$, leading to different source and relay data rates $R_S$ and $R_R$. The total bus data rate per WDM channel is $R_S+R_R=250$\,Gbit/s. We can see that strategies b) and c) have similar performance, especially when $w=3$, and hence strategy b) is a good candidate for use with the OE protocol. The proposed method can also be used to trade throughput versus distance in networks.

\vspace*{-1ex}
\section{Conclusions}
In this paper, we have analyzed a simple regenerative network and have experimentally shown reach increases by a relay. The latter additionally enables networking functionalities which maximize the total network distance by exploiting the unique properties of SC-LDPC codes.

\bibliographystyle{abbrv}
\begin{spacing}{0.75}
\setlength{\bibsep}{2pt}

\end{spacing}
\vspace{-4mm}

\end{document}